# Probing Charge Dynamics in Amorphous Oxide Semiconductors by Time-of-flight Microwave Impedance Microscopy

Jia Yu[1], Yuchen Zhou[2], Xiao Wang[2], Xuejian Ma[1], Ananth Dodabalapur*[2], Keji Lai*[1]

[1] Department of Physics, University of Texas at Austin, Austin TX 78712, USA

[2] Department of Electrical and Computer Engineering, University of Texas at Austin, Austin TX 78712, USA

* E-mails: ananth.dodabalapur@engr.utexas.edu, kejilai@physics.utexas.edu

## Abstract

The unique electronic properties of amorphous indium gallium zinc oxide (a-IGZO) thin films are closely associated with the complex charge dynamics of the materials. Conventional studies of charge transport in a-IGZO usually involve steady-state or transient measurements on field-effect transistors. Here, we employed microwave impedance microscopy to carry out position-dependent time-of-flight (TOF) experiments on a-IGZO devices, which offer spatial and temporal information of the underlying transport dynamics. The drift mobility calculated from the delay time between carrier injection and onset of TOF response is 2 ~ 3 $cm^2/V \cdot s$, consistent with the field-effect mobility from device measurements. The spatiotemporal conductivity data can be nicely fitted to a two-step function, corresponding to two coexisting mechanisms with a typical timescale of milliseconds. The competition between multiple-trap-and-release conduction through band-tail states and hopping conduction through deep trap states is evident from the fitting parameters. The underlying length scale and time scale of charge dynamics in a-IGZO are of fundamental importance for transparent and flexible nanoelectronics and optoelectronics, as well as emerging back-end-of-line applications.

**Introduction**

Amorphous oxide semiconductors (AOSs) are advanced electronic materials with good carrier mobility, large-area uniformity, and low manufacturing cost.[1-3] Of special interest in the AOS family is amorphous indium gallium zinc oxide (a-IGZO), which exhibits more than ten times higher mobility than hydrogenated amorphous silicon (a-Si:H). Since the first successful demonstration in 2004,[2] a-IGZO thin-film transistors (TFTs) have been extensively investigated for applications in flexible electronics, flat-panel displays, and various types of sensors.[3] As a result, the study of charge dynamics in a-IGZO is not only scientifically significant for the understanding of disordered solids but also technologically relevant for transparent and flexible nanoelectronics and optoelectronics.

The remarkable properties of a-IGZO stem from certain unique features in its electronic structure. Similar to other amorphous semiconductors, there exist abundant trap states inside the energy gap due to the lack of long-range periodicity. Transport across strongly localized states exhibits the usual hopping behavior. On the other hand, owing to the highly overlapping spherical $s$ orbitals of heavy metal cations in a-IGZO, the edges of conduction and valence bands, although not as sharp as those in crystalline semiconductors, are reasonably well defined with shallow band tails.[4] Charge transport through these band-tail states is much faster than those in typical organic semiconductors, which can be described by the multiple trap and release (MTR) model.[5,6] These two transport mechanisms are not mutually exclusive, and whichever dominates in a particular device depends on the sample growth condition and the operating temperature.[7,8] The competition between thermally activated hopping and MTR conduction is a defining feature of a-IGZO, leading to its intermediate properties between highly ordered and highly disordered systems.[5,6,9,10]

Traditional studies of charge transport in semiconducting materials are carried out in the configuration of field-effect transistors (FETs). For a-IGZO thin films, however, the extracted filed-effect mobility is only a measure of the averaged behavior across the source and drain electrodes. In contrast, transient time-of-flight (TOF) measurements, which detect the temporal evolution from carrier injection at the source to subsequent extraction at the drain, can directly resolve the drift motion of mobile charges in the time domain.[11-17] To date, TOF studies are mostly performed on TFT devices with fixed source and drain electrodes. In this work, we report the results of position-dependent transient experiments on a-IGZO devices by microwave impedance microscopy (MIM).[18,19] The time interval between the rising edge of a step-like bias excitation and the incipience of MIM response scales with the square of the tip-electrode distance. The drift mobility calculated from this delay time is consistent with the field-effect mobility. More importantly, the buildup of MIM signals at different locations from the source electrode can be fitted to an exponential function plus an error function, which presumably correspond to contributions from electrons in the band-tail states and deep-trap states, respectively. The competition between the MTR and hopping mechanisms is evident from the extracted parameters in the curve fitting. Our work reveals the underlying length scale and time scale of charge dynamics in a-IGZO, which are of fundamental importance to their device applications.

**Results**

The 20-nm-think a-IGZO thin films in this experiment are deposited onto heavily doped n-type Si substrates with 90 nm thermal $SiO_2$ by RF-sputtering. The nominal composition of $Ga_2O_3$:$In_2O_3$:$ZnO$ is in a ratio of 1:2:2 (sputtering target purchased from Kurt J. Lesker Company). The deposition is performed in Ar gas with 7% $O_2$ content and a total pressure of 5 mTorr. The films are then annealed on a hot plate for 1 h at 400 °C to introduce oxygen vacancies. For TFT devices, both the a-IGZO film and the Al contact electrodes are directly deposited through shadow masks. Since neither lithography nor etching is involved in the process, any extrinsic effect due to sample fabrication is minimized. Transport measurements were carried out in the atmosphere using a Keysight 4155C Semiconductor Parameter Analyzer in the dark environment. All experiments are performed at room temperature.

Figure 1a shows the schematic of our steady-state experiment to calibrate the MIM response on a back-gated a-IGZO TFT device. Here, the 1 GHz microwave excitation signal is delivered to the cantilever probe by an impedance-match network. Since the dimension of the tip apex (~ 100 nm) is much smaller than the wavelength of the microwave (30 cm), the near-field tip-sample interaction can be described by the lumped-element model.[18] The small variation of the tip impedance during the scanning is detected by MIM electronics. The MIM outputs are proportional to the imaginary (MIM-Im) and real (MIM-Re) parts of the tip-sample admittance, from which the local microwave conductivity can be extracted.[18,19] In Figure 1b, when the transistor turns on, the transfer curves show little gate hysteresis between forward and backward sweeps, indicative of an insignificant amount of interface charges between the gate dielectric and the channel material. In contrast, the off-state current after the backward sweep is two orders of magnitude higher than that before the forward sweep, presumably due to the presence of trap carriers that are not fully depleted. Using the method in Reference 20, the threshold voltage of this device is determined as $V_{th}$ = 2.13 V. Figure 1c displays a number of 1 GHz MIM images across the boundary between the Al contact and a-IGZO film at various back-gate voltages (complete data in Supplementary Information Section S1). The MIM is a moveable probe that detects the electrical conductivity underneath the tip,[18] rather than a fixed electrode that measures the drain current in conventional TOF experiments. Quantitative analysis of the MIM data by finite-element modeling[19] is identical to that in our previous report,[21] which is included in Supplementary Information Section S2. In Figure 1d, we plot the measured and simulated MIM-Im signals against the extracted local conductivity $\sigma$ of the a-IGZO thin film. In the context of this work, we will utilize the linear MIM-Im vs. $\sigma$ relationship when the MIM-Im signal is below 1 V such that we may directly present the raw MIM data without converting them to conductivity values.

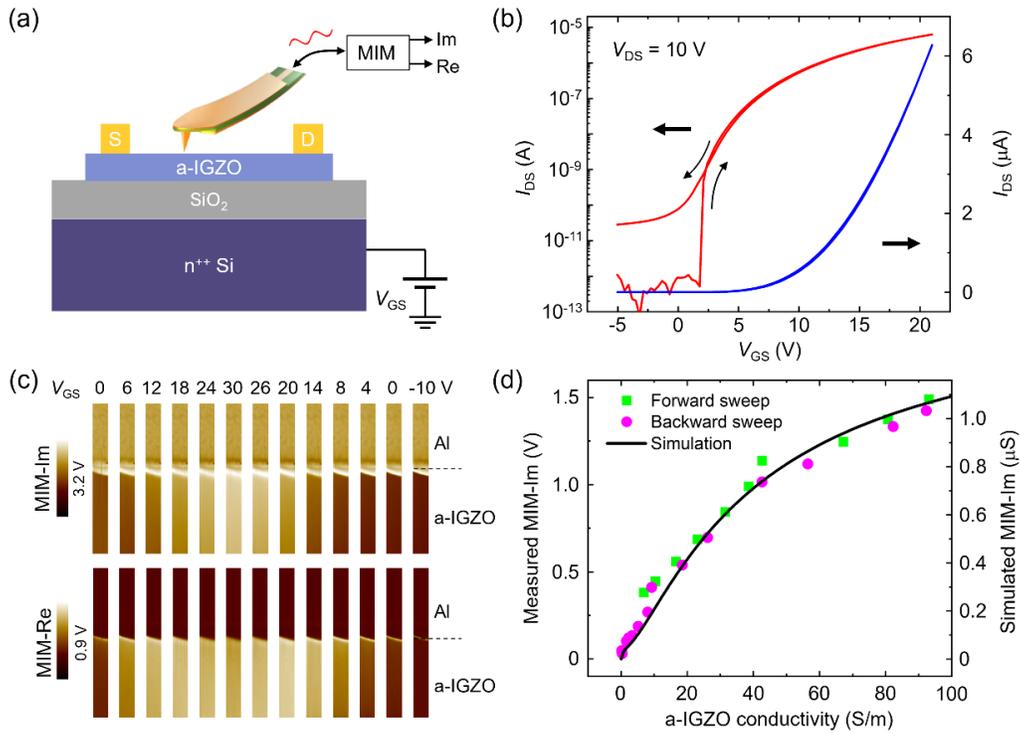

**Figure 1.** (a) Schematic of MIM imaging on a back-gated a-IGZO TFT device. (b) Transfer characteristics of an a-IGZO transistor at $V_{DS}$ = 10 V and a sweep rate of 0.76 V s$^{-1}$ in both logarithmic and linear scales. (c) Selected MIM images across the Al/a-IGZO boundary as $V_{GS}$ increases from 0 V to 30 V and then decreases to -10 V. The size of each image is 55 μm × 5.6 μm. (d) Measured and simulated MIM-Im signals as a function of the a-IGZO conductivity, with a roughly linear relation when the MIM-Im signal is below 1 V.

The experimental setup of our TOF measurement is illustrated in Figure 2a, where the source electrode is grounded and a positive step-like voltage $\Delta V_{GS}$ from a function generator is applied on the Si back gate. The potential of the insulating a-IGZO channel is pulled to $V_G$ at $t = 0$, resulting in a local lateral electrical field that drives the drift current.[22] As electrons arrive at a particular point, the local conductivity increases, and the local potential gradually equilibrates with the grounded source. The high carrier concentration at the source also leads to a diffusion current into the channel. However, since $V_{GS} - V_{th} \gg k_B T/q = 26$ meV in our present case, the diffusion current is overshadowed by the drift current. In reality, the configuration in Figure 2a is achieved by using a low-frequency ($< 100$ Hz) square wave and only focusing on the rising edge. Note that the falling edge is not a simple reversal of the TOF process and will not be discussed here. Again, our experiment differs from conventional TOF measurements in that the MIM is a nanoscale conductivity probe. The MIM response at various distance $L$ from the electrode (image in Figure 2a, complete device image in Supplementary Information Section S3) is recorded by a high-speed oscilloscope. The ultimate temporal resolution (1 ~ 10 ns) of time-resolved MIM is limited by reciprocal of the microwave frequency. In this experiment, the temporal resolution is determined by the finite rise time of ~ 100 ns of the function generator (Supplementary Information Section S4). However, it will become obvious that the relevant time scale here is in the order of ms. As a result, we have used an amplifier with a low-pass filter to improve the signal-to-noise ratio such that the measurement can be performed in a single-shot manner. As shown in Supplementary Information Section S4, we also confirm that the time-resolved MIM signals are not distorted by the filtering.

Before presenting our position-dependent time-resolved MIM data, it is instructive to first review the analytical description of the TOF method for trap-free band transport. Adapting the study on Si transistors, the problem is formulated by two equivalent models[23] – transmission-line approach[24] and quasi-one-dimensional approach – and obtained the differential equation for the channel potential $V_{ch}(x, t)$ at a distance $x$ from the source electrode and time $t$ as,

$$\mu \frac{\partial}{\partial x}\left(V' \frac{\partial V'}{\partial x}\right) = \frac{\partial V'}{\partial t} \qquad (1)$$

, where the substitution $V' = V_{GS} - V_{th} - V_{ch}(x, t)$ and $\mu$ is the drift mobility.[23] The carrier density $n(x, t)$ can then be calculated by Poisson's equation under the long-channel approximation,

$$\frac{qn(x,t)}{\varepsilon_0\varepsilon_r} = E_y = \frac{V_{GS} - V_{th} - V_{ch}(x,t)}{t_{ox}} \propto V' \tag{2}$$

, where $E_y$ is the vertical electric field in SiO$_2$, $q$ is the elemental charge, $\varepsilon_r$ and $t_{ox}$ are the relative dielectric constant and thickness of SiO$_2$, respectively. The differential Equation (1) cannot be solved analytically. Using the boundary conditions of $V'(0,t) = 0$ and $V'(x,0) = V_{GS} - V_{th}$, one can numerically evaluate the delay time $\tau$ between the carrier injection and onset of rise in $V_{ch}$ as,

$$\tau(L) \approx 0.38 \frac{L^2}{\mu(V_{GS} - V_{th})} \tag{3}$$

After this initial transit time, charges start to build up at position $x = L$ and asymptotically approaches the saturation value $n_0$ following the numerical solution,

$$n(L,t) \approx n_0\left(1 - e^{-(t-\tau)/T_0}\right) \tag{4}$$

, where $T_0$ is the characteristic time of the charging process. In this experiment, the MIM detects the local conductivity $\sigma(x,t)$, which is proportional to $n(x,t)$ as the drift mobility in this a-IGZO sample is roughly a constant (to be justified later).

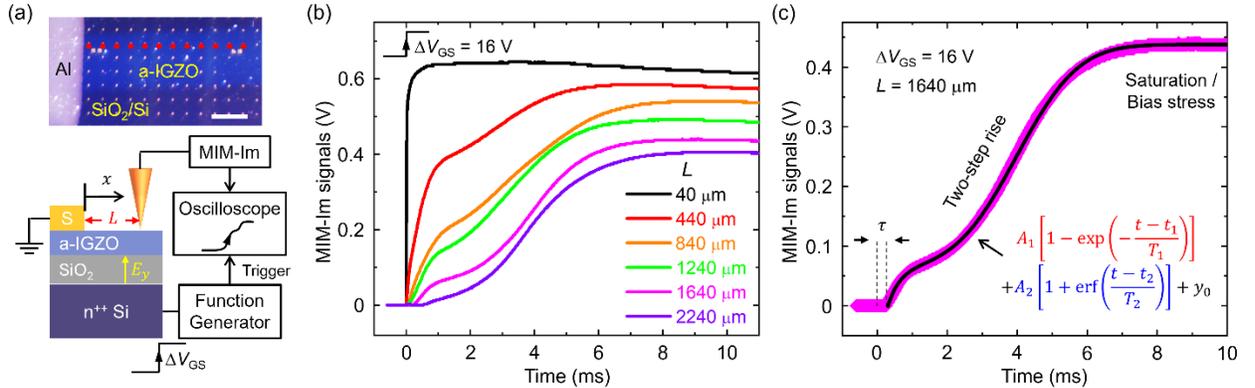

**Figure 2.** **(a)** Schematic of the time-of-flight MIM setup. The function generator sends a step-like bias voltage $\Delta V_{GS}$ to the Si back gate and triggers the oscilloscope that measures the MIM-Im signals. The tip is positioned at various distance $L$ away from the Al electrode, as marked by red dots. Scale bar: 500 μm. **(b)** Time-dependent MIM-Im signals under $\Delta V_{GS} = 16$ V at various positions. **(c)** Highlighted time evolution at $L = 1640$ μm, showing the delay time $\tau$, two-step rise that can be fitted to the function indicated in the plot, and signal saturation and bias-stress effect.

Figure 2b shows the time-dependent MIM-Im signals under $\Delta V_{GS} = 16$ V at various locations (full set of raw data in Supplementary Information Section S5). We should point out that there exists a sudden (below our temporal resolution) jump of the MIM-Im signal at $t = 0$, which occurs even at locations millimeters away from the source. Such an instantaneous response is due to the vertical field effect between the back gate and the metal tip, which induces a small amount of carriers underneath the probe, analogous to the displacement current effect in conventional TOF measurements.[23] Since the jump is not related to the lateral charge transport, we have removed it from the subsequent data analysis. Figure 2c highlights the complete time evolution of MIM-Im signals at $\Delta V_{GS} = 16$ V and $L = 1640$ μm, showing the delay time $\tau$ due to the motion of charges in the channel, the buildup of electrons that can be fitted to two-step rises, and the saturation of signals and small decay afterwards caused by the bias-stress effect.[25, 26] In the following, we will elaborate on individual steps and discuss the physics behind each process.

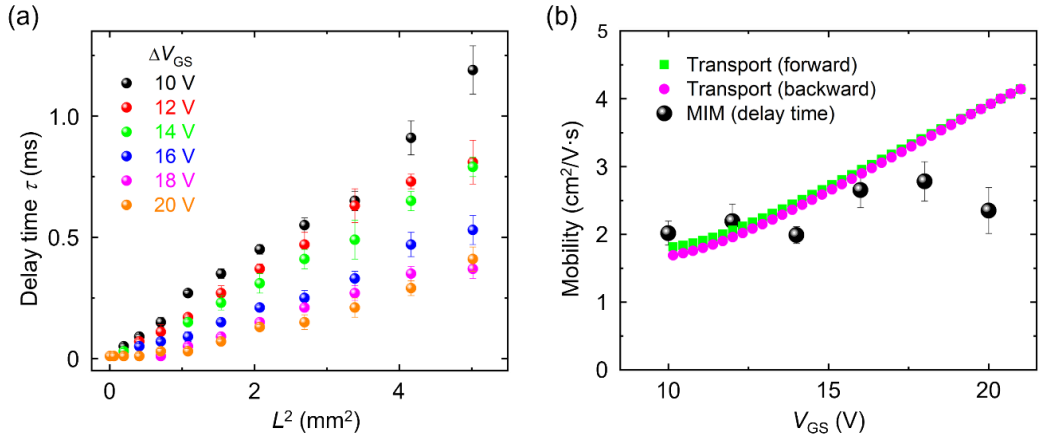

**Figure 3.** (a) $\tau - L^2$ relation under various bias excitations. (b) Comparison between drift mobility extracted from the delay time analysis and average field-effect mobility extracted from transport data.

Figure 3a plots the delay time $\tau$ as a function of $L^2$ under various bias excitations. The linear $\tau - L^2$ relation suggests that the traditional TOF analysis is still applicable for the fastest carriers arriving at the detection point.[27] Using Equation (3), we can extract the drift mobility from slopes of the linear fits, as plotted in Figure 3b. The result is also consistent with the average field-effect mobility calculated from the steady-state transport data[20] in Figure 1b. We therefore conclude that the mobility value of 2 ~ 3 cm$^2$/V·s provides a good measure of transport properties through the MTR mechanism at the shortest time scale in this a-IGZO sample. Note that recent efforts[28-30] have improved the mobility of IGZO thin films to values above 10 cm$^2$/V·s. In this study, we are focusing on low-temperature and low-cost recipes that are compatible with amorphous semiconductor processing, where a mobility of 1 ~ 10 cm$^2$/V·s is sufficient for TFT applications. As seen below, this intermediate mobility range allows us to access the regime where multiple transport mechanisms compete with each other.

We now discuss the increase in MIM-Im signals after the initial delay, corresponding to the buildup of charges and thus electrical conductivity in the channel. From Figure 2b, it is obvious that the process cannot be described by Equation (4) as a single exponential function, and a two-term function is necessary for data fitting. In fact, while the first rising step follows the exponential characteristics, the line shape of the second step is much more gradual with an inflection point in the middle. As suggested in Figure 2c, the data can be fitted to an empirical expression as,

$$y(t) = A_1 \left[1 - \exp\left(-\frac{t - t_1}{T_1}\right)\right] + A_2 \left[1 + \mathrm{erf}\left(\frac{t - t_2}{T_2}\right)\right] + y_0 \quad (5)$$

, with the fitting parameters of $A_1(A_2)$: amplitude of the exponential (error) function; $T_1(T_2)$: time constant of the exponential (error) function; $t_1$: starting point of the exponential function; $t_2$: midpoint of the error function; $y_0$: constant to match the saturation value. As shown in Supplementary Information Section S6, the time-resolved MIM-Im data at every $L$ and under each $\Delta V_{GS}$ can be satisfactorily fitted to Equation (5). We also plot all fitting parameters and provide discussions in Supplementary Information Section S7. Last but not least, the fitting does not cover the slow decay of signals after saturation, which is well understood as the bias stress effect.[25,26]

Figure 4a shows the position dependence of $A_1$ and $A_2$ under several $\Delta V_{GS}$'s. Close to the contact, the MIM response is dominated by the exponential function ($A_1$) term, which rapidly decreases around $L = 1$ mm and vanishes at $L = 2$ mm and beyond. In contrast, the error function ($A_2$) term is relatively small close to the source electrode, and starts to rise at $L \sim 1$ mm and saturates at $L \sim 2$ mm. Finally, a comparison between $T_1$ and $T_2$ is also informative. As seen in Figure 4b and 4c, except for the point closest to the electrode, both time constants are essentially independent of the tip position. Both $T_1$ and $T_2$ are decreasing with increasing gate bias, presumably due to the rapid filling of empty states under high bias. On the other hand, the bias-independent ratio of $T_1/T_2 \sim 0.2$ (Supplementary Information Section S7) indicates that $\Delta V_{GS}$ changes both time constants simultaneously. In summary, the buildup of charges in the a-IGZO channel is a two-step process, with the exponential function term dominating at the shorter time and length scales and the error function term dominating at the longer time and length scales.

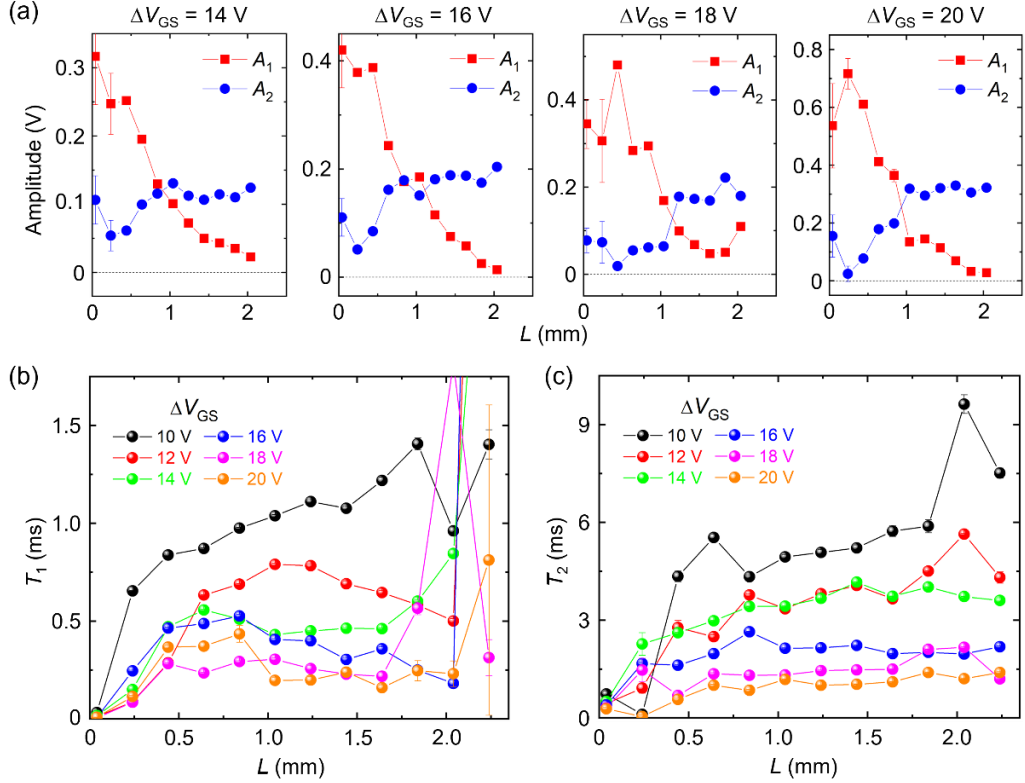

**Figure 4.** (a) Position dependence of $A_1$ and $A_2$ under various gate biases. (b) Position dependence of $T_1$ and (c) $T_2$ under various gate biases. The values are essentially independent of the tip position except for the first 1 or 2 points. Note that $T_2$ is generally greater than $T_1$ by a factor of ~ 5.

## Discussions

The spatially and temporally resolved MIM data presented above vividly demonstrate the complex charge dynamics in a-IGZO thin films. While a rigorous TOF analysis with two transport mechanisms is beyond the scope of this work, it is possible to develop a phenomenological model that captures the key physics here. A key feature of MTR transport is that carriers thermalize to extended states in the conduction band where they can move by a combination of band transport and hopping transport until they are trapped again.[31] The small energy difference between tail states and the conduction band makes such transport likely at short time scales. Thus, the traditional TOF analysis of delay time (Equation 3) and exponential charging (Equation 4) still applies to the fastest moving carriers in the channel, as indicated by red arrows in Figure 5. When propagating along the shallow states, however, a portion of the electrons may fall into deep traps (black arrows in Figure 5) and hardly return to the band tail.[9] Although we do not know *a priori* the analytical result of TOF signals for hopping transport, it is reasonable to assume that the rise of MIM signals takes the form of an error function that is the integral of a Gaussian peak.[17, 32] In other words, the time-resolved data will gradually transition from an exponential function (the $A_1$ term) at short distance to an error function (the $A_2$ term) at long distance, consistent with the experimental observation. Moreover, since hopping through deep traps is slower than MTR transport through band-tail states, the corresponding time constant of the former ($T_2$) is longer than the latter ($T_1$), again consistent with our TOF data. With the rising Fermi level under higher gate biases, the process of band filling is accelerated, leading to smaller $T_1$ and $T_2$. Finally, trapping of carriers by deep in-gap states will continue to occur even after the error-function term saturates, resulting in the gradual decrease of MIM signals (bias stress effect) at a time scale of $t \sim 10$ ms. We caution that the analysis above does not explain the very short time constant (still above our temporal resolution) within 50 μm from the source electrode, which may be related to contact resistance due to Schottky contact between Al and a-IGZO. Future theoretical and experimental studies are needed to understand the spatiotemporal MIM response in short-channel devices. Given the intermediate mobility in this work where both MTR and hopping mechanisms are equally important, another interesting future direction is to investigate samples with a wide range of mobility and device performance. For instance, in devices with similar dimensions but higher mobilities, even though the carriers will still relax to lower energies with time, the MTR mechanism may dominate over hopping mechanism, and the latter is expected to take place at a longer distance from the electrode.

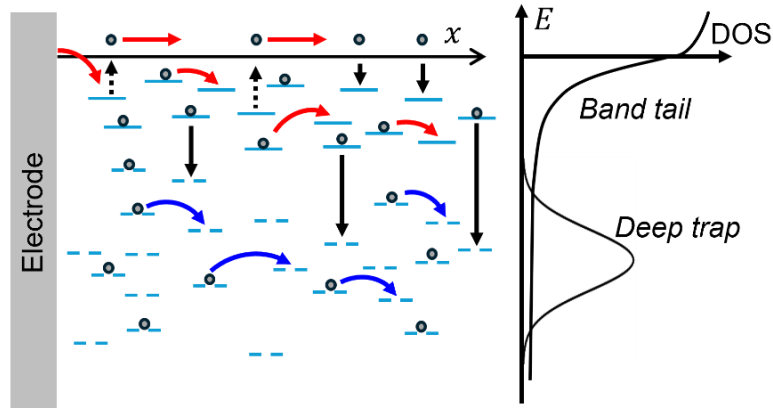

**Figure 5.** Left: Schematic of distribution of in-gap states in a-IGZO with a source electrode. Electrons are injected from the contact and drift to the right. Red and blue arrows represent the transport through band-tail (including band transport) and deep-trap states, respectively. Black dashed arrows represent thermal activation of trap carriers to the conduction band. Black solid arrows represent the trapping of electrons into deeper states. Right: Density of states in a-IGZO, consisting of an exponential band tail and a Gaussian-like localized band.

**Conclusion**

In summary, we report the results of position-dependent time-of-flight experiment on a-IGZO thin films by microwave impedance microscopy. By sending a step-like voltage excitation to the gate electrode and monitoring the transient signals, we are able to resolve the delay time between the carrier injection and the onset of MIM response, the buildup of electrical conductivity in the channel, and the gradual loss of carriers after signal saturation due to the bias-stress effect. Mobility extracted from the TOF analysis is about 2 ~ 3 cm$^2$/V·s. In particular, we show that the multiple trap-and-release transport through band-tail states with a characteristic time constant around 1 ms is responsible for the initial transit time and the rise of MIM signals close to the source electrode. As electrons are trapped into strongly localized states at longer time and length scales, hopping transport through deep traps with a characteristic time constant of a few ms gradually takes over the charge dynamics, resulting in error-function-like TOF characteristics and the bias-stress effect. Our work contributes to the understanding of competing transport mechanisms in amorphous oxide semiconductors, which are important for their applications in nanoelectronics and optoelectronics.

**Experimental Methods**

**TOF Microwave Impedance Microscopy measurements.** The MIM experiments in this work were performed on an AFM platform (ParkAFM XE-70). The customized shielded cantilevers are commercially available from PrimeNano Inc. For the TOF measurement, a square wave voltage (40 or 75 Hz, 10 to 20 V) was generated from Hewlett Packard 214B pulse generator. The voltage was connected to the Si back gate and sent to CH1 of the oscilloscope (DS6062, RIGOL Technologies USA Inc.) as the reference signal. The MIM-Im signal, passing through either a low-f amplifier (54 dB, 20 kHz) or a fast amplifier (28 dB, 350 MHz, SR445A, Stanford Research System Inc.) after the mixer, was sent to CH2 of the oscilloscope. We could thus collect the gate voltage and MIM-Im signals simultaneously.

**Supporting Information**

Supporting Information is available online, including a complete set of gate-dependent MIM images, quantitative analysis of MIM signals, complete device image, characterization of time-of-flight electronics, full position-dependent tr-MIM data under $\Delta V_{GS} = 16$ V, complete spatiotemporal MIM data, and curve fitting with fitting parameters.

**Acknowledgements**

The MIM work was supported by the U.S. Department of Energy (DOE), Office of Science, Basic Energy Sciences, under Award DE-SC0019025. A.D. acknowledges support from the Semiconductor Research Corporation (SRC) Task ID #2962.001, the National Science Foundation under Grant NNCI-2025227, and the Keck Foundation under Grant 26753419. K.L. acknowledges support from the Welch Foundation under Grant F-1814. This work was partly done at the Texas Nanofabrication Facility supported by NSF Grant NNCI-1542159.

**Author Contributions**

K.L. and A.D. conceived the project. Y.Z. and W.X. fabricated the devices and performed the transport measurements. J.Y. tested TOF setup with X.M., collected data and performed analysis with Y.Z., and drafted the manuscript with K.L. All authors have contributed to the manuscript and given approval to the final version of the manuscript.

**Conflict of Interest**

The authors declare no competing financial interests.

**Data Availability Statement**

The data that support the findings of this study are available from the corresponding author upon reasonable request.